\documentclass[aip,numerical,reprint,amssymb,secnumarabic,superscriptaddress,twocolumn]{revtex4-1} 

\usepackage{graphicx} 
\usepackage{natbib} 
\usepackage[usenames,dvipsnames]{color} 
\usepackage{amsmath} 
\usepackage[urlcolor=blue, hyperindex, colorlinks, bookmarks=true,linkcolor=black,citecolor=blue]{hyperref} 
\usepackage{amssymb} 
\usepackage{soul} 
\usepackage{ifthen} 
\usepackage{upgreek}

\def\be#1\ee{\begin{equation}#1\end{equation}}
\def\ba#1\ea{\begin{align}#1\end{align}}
\def\bg#1\eg{\begin{gather}#1\end{gather}}

\def\shownote{1} 
\newcommand{\note}[1]{\ifthenelse{\shownote=1}{\textcolor{Red}{[[#1]]}}{}}
\def\showaddmat{1} 
\newcommand{\addmat}[1]{\ifthenelse{\showaddmat=1}{\textcolor{Gray}{[[#1]]}}{}}

\begin{document}


\title[]{Solving the Jitter Problem in Microwave Compressed Ultrafast Electron Diffraction Instruments: Robust Sub-50~fs Cavity-Laser Phase Stabilization }

\author{M. R. Otto}
\email{martin.otto@mail.mcgill.ca}|
 \affiliation{
Department of Physics, Center for the Physics of Materials, McGill University, 3600 University Street, Montreal, QC, CA 
}%
\author{L. P. Ren\'e de Cotret}
 \affiliation{
Department of Physics, Center for the Physics of Materials, McGill University, 3600 University Street, Montreal, QC, CA 
}%
\author{M. J. Stern}
 \affiliation{
Department of Physics, Center for the Physics of Materials, McGill University, 3600 University Street, Montreal, QC, CA 
}%
\author{B. J. Siwick}%
 \email{bradley.siwick@mcgill.ca}
 \affiliation{
Department of Physics, Center for the Physics of Materials, McGill University, 3600 University Street, Montreal, QC, CA 
}%
\affiliation{
Department of Chemistry, McGill University, 801 Sherbrooke Street W, Montreal, QC, CA 
}%

\date{\today}

\begin{abstract}
We demonstrate the compression of electron pulses in a high-brightness ultrafast electron diffraction (UED) instrument using phase-locked microwave signals directly generated from a mode-locked femtosecond oscillator. Additionally, a continuous-wave phase stabilization system that accurately corrects for phase fluctuations arising in the compression cavity from both power amplification and thermal drift induced detuning was designed and implemented. An improvement in the microwave timing stability from 100~fs to 5~fs RMS is measured electronically and the long-term arrival time stability ($>~$10 hours) of the electron pulses improves to below our measurement resolution of 50~fs. These results demonstrate sub-relativistic ultrafast electron diffraction with compressed pulses that is no longer limited by laser-microwave synchronization.
\end{abstract}

\maketitle

Ultrashort electron pulses are finding diverse applications in research aimed at imaging the dynamic structure of matter~\cite{UEDRepProgPhysics2011,Morrison2014,Miller2014,Schwoerer2016}. Generation of these pulses normally starts with photoemission driven by a femtosecond laser pulse at a photocathode, after which Coulomb repulsion internal to the photo-generated bunch takes hold (space-charge) broadening both the temporal duration and energy distribution~\cite{Siwick2002}. Unmodified by external fields, these space-charge dynamics result in a trade-off between pulse fluence and time resolution that is detrimental to ultrafast electron diffraction and imaging experiments.  As a result, there have been a number of efforts to correct such broadening through the addition of electron pulse compression strategies that employ microwave~\cite{TvOpulsecompression2007,ChatelainAPL2012,Gao2012,vanOudheusdenPRL2010,GliserinNJP2012,Zandi2017}, terahertz~\cite{Kealhofer2016} and DC electric fields~\cite{Mankos2017,Wang2012,Kassier2009}. These approaches work by inverting the space-charge driven expansion that occurs naturally in the pulse, leading to a temporal focus downstream from the pulse-field interaction.  Microwave compression in particular has been demonstrated to be very effective in the single shot limit, yielding electron pulses below 100 fs at 100 keV~\cite{vanOudheusdenPRL2010} and very recently sub-10~fs at 7~MeV~\cite{Maxson2017} that contain more than $10^5$ electrons.  Unfortunately, the stability of the cavity-laser synchronization systems that have been employed to date have been insufficient to provide pulse duration limited time-resolution in ultrafast electron diffraction instruments over longer data acquisition times (several hours).  Published reports have all concluded that time-resolution in microwave compressed instruments has been closer to 400~fs due to ``time-zero" drift that results from various cavity-laser phase syncrhonization instabilities~\cite{ChatelainAPL2012,Gao2012,Zandi2017} that are evident in the frequency range from kHz to $\upmu$Hz.  As a result of these drifts, the primary benefit of microwave pulse compression to date and been an increase in bunch charge rather than a dramatic improvement in time resolution.

For UED pump-probe experiments, synchronizing the laser system with a microwave signal has been previously achieved by phase-locking loops (PLL) using external voltage-controlled oscillators~\cite{Kiewiet2002,ChatelainAPL2012,Gao2012}, or repetition rate multiplication techniques~\cite{Gliserin2013} which involve optical enhancement cavities.  Both approaches involve the derivation of a locked harmonic in the GHz range with sufficient spectral power and sufficiently \it low \rm phase noise.  Timing and frequency stability is fundamentally limited by amplitude-phase conversion inherent to the photodetection process~\cite{Ivanov2003,Taylor2011,Zhang2012} and depends on the pulse energy stability of the laser.  For the case of the phase-locked loop, amplitude-phase conversion also manifests in microwave mixers when comparing two signals, yielding phase errors produced by power fluctuations.  For UED synchronization systems to date, the focus has been primarily on the frequency range above 1~Hz, with minimal consideration of drift on timescales up to several hours, which are of particular relevance for experiments.  Such drifts cause $t=0$ to change over the course of many pump-probe delay scans and arise most significantly in sensitive elements such as the compression cavity and power amplification which are typically omitted from the synchronization configuration~\cite{Wallbran2015}.  In this work, we demonstrate stable, passive generation of a $3~$GHz signal by direct photodetection of the laser pulse train and its use for compression of electron pulses in a 100~keV in a high-brightness ($10^6~e^-/$pulse) ultrafast electron diffraction instrument.  We also present an all-microwave active synchronization enhancement system which measures and compensates for phase fluctuations arising in the compression system. We measure greater than a ten-fold improvement in laser-microwave synchronization quality by directly measuring phase changes of the cavity field using an integrated antenna.  We further demonstrate, using a streak camera, that the phase stabilization system improves the long-term stability of the pulse arrival time and the temporal impulse response function of the UED instrument by correcting for phase over a very broad low-frequency band. This improved performance is due to the elimination of several sources of phase instability inherent in previous approaches; i) amplitude phase errors in PLL generation of microwave signal, ii) Phase instability in power amplification iii) Phase drift in the cavity response due to thermal induced frequency detuning.
\begin{figure}
  \centering
    \includegraphics[width=8.5cm]{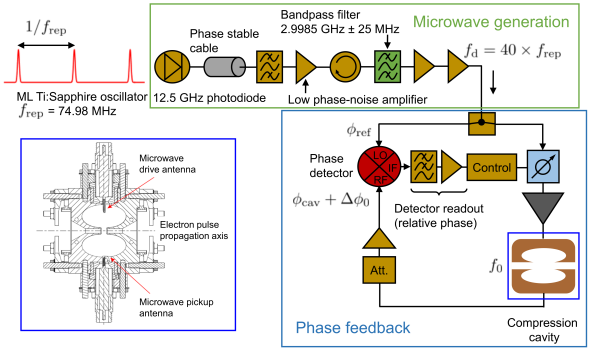}
  \caption{Direct generation of synchronized microwaves and active phase stabilization for electron pulse compression.  A portion of the output of a mode-locked Ti:Sapphire oscillator ($f_{\textup{rep}}\approx~$74.98~MHz) is incident on a high-bandwidth ($12.5~$GHz) GaAs photodiode.  The 40$^{\textup{th}}$ harmonic is generated by filtering using narrow bandpass filters and ultra-low phase noise microwave amplifiers.  The power of the synchronized drive signal $f_{\textup{d}}=40\times f_{\textup{rep}}\approx 2.9985~$GHz is roughly $+16~$dBm after passing through the final amplifier operated in saturation. The signal is split and the first serves as a reference signal carrying a phase  $\phi_{\textup{ref}}$ and is fed to the local oscillator (LO) port of a phase detector.  The second signal is amplified by a CW solid-state power amplifier and drives the resonant cavity for electron pulse compression.  An antenna integrated into the cavity retreives the phase $\phi_{\textup{cav}}+\Delta\phi_0$ by coupling directly to the electric field and produces an output signal which is attenuated and amplified to reach a power identical to that of the reference signal before coupling to the second input (RF) of the phase detector.  The feedback system maintains a constant value of $\phi_{\textup{ref}}-\phi_{\textup{cav}}$ by shifting the phase of $\phi_{\textup{cav}}$ in order to eliminate fluctuations $\Delta\phi_0$.}\label{fig:feedback_cavity}
\end{figure}

The master clock of the instrument used in this work is a mode-locked Ti:Sapphire oscillator with a fundamental repetition rate of $f_{\textup{rep}}\simeq 74.98~$MHz. A synchronized microwave signal is generated by sampling a portion of the oscillator laser output using a fast photo-diode (Newport 818-BB-45) with a bandwidth of 12.5~GHz (Fig.~\ref{fig:feedback_cavity}).  This signal is filtered by a coarse band-pass filter (Mini-Circuits VBF-2900+) to select the relevant harmonic ($f_{\textup{d}}=40\times f_{\textup{rep}}$) and sent to an ultra-low phase noise narrow band amplifier (Miteq AMF-2F-LPN) providing roughly 30 dB of gain.  The amplified output filtered by a cavity band-pass filter centered at 2.9985~GHz with a bandwidth of 50~MHz to further isolate the desired harmonic signal. Reflected signals from harmonics outside the pass-band are isolated and dumped to a 50~$\Omega$ load to maintain directivity of the signal generation circuit.  The signal is amplified by a low phase noise amplifier (Holzworth HX2400) and then a saturated low noise amplifier (Fairview SLNA-060-40-09) in order to achieve optimal power stability.  The signal is then split into two paths.  The first path is continuous-wave amplified to high-power (typically 46-48~dBm) and used to drive the microwave electron pulse compression cavity (see Fig.~\ref{fig:feedback_cavity}) which has a TM$_{010}$ resonant mode of $f_0=2.9985~$GHz at 19~$^{\textup{o}}$C, an unloaded quality factor of $Q\simeq 1.2\times 10^4$ and a bandwidth of $\delta\approx 250~$kHz.  In order to improve the synchronization quality of the instrument, we must account for phase changes produced by elements in this first signal path relative to those in the second, which we will consider to be our stable phase reference.
\begin{figure}
  \centering
    \includegraphics[width=8.5cm]{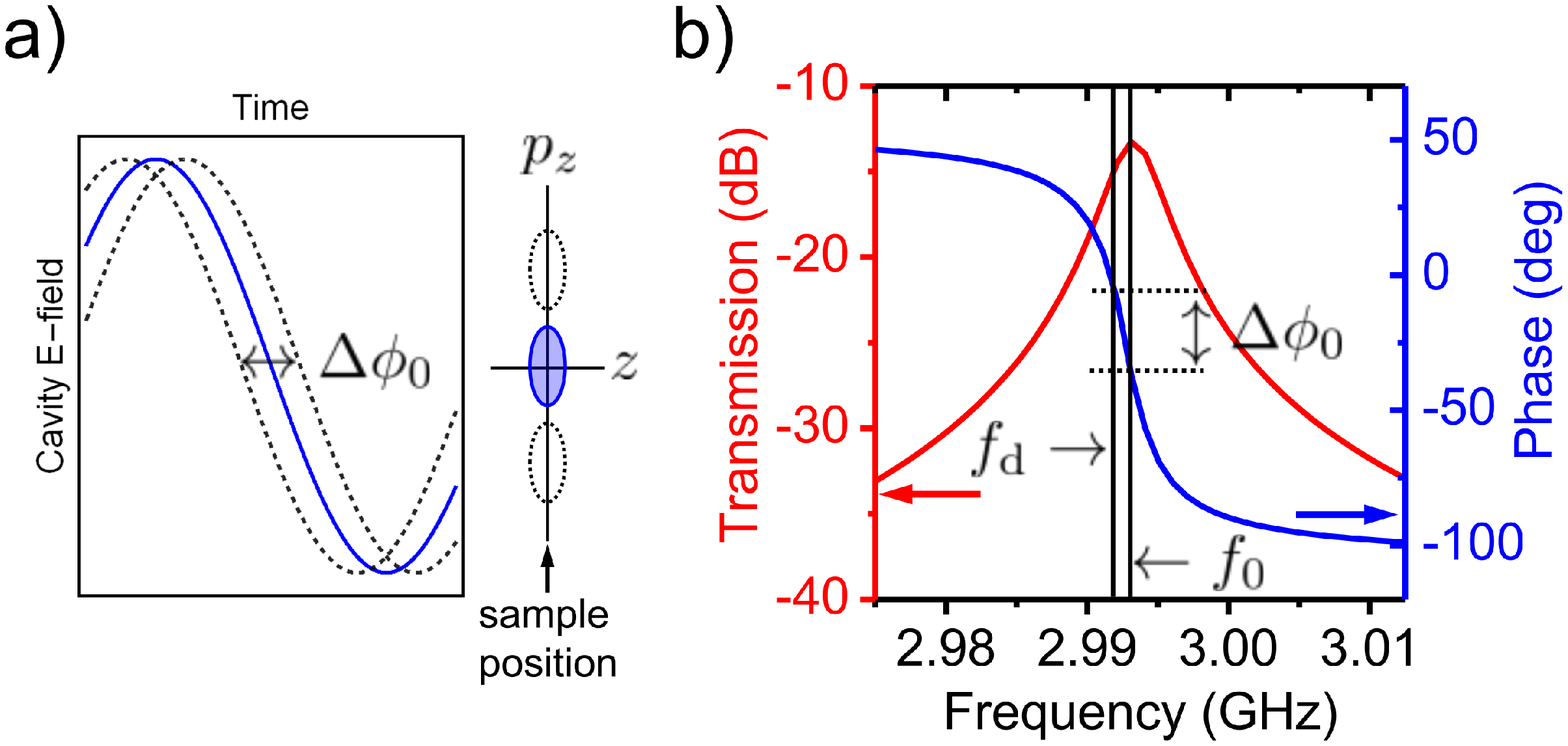}
  \caption{a)  Illustration of the effect of phase instabilities. Changes in phase $\Delta\phi_0$ of the electric field oscillation in the compression cavity relative to the arrival of an electron pulse cause a change in momentum transfer to the compressed pulse $\Delta p_z$. b) Simulation of the amplitude and phase of the $S_{21}(f)$ transmission function for our $\sim 3~$GHz microwave compression cavity. Detuning between the drive frequency $f_0$ and cavity resonance $f_{\textup{d}}$ lead to a phase shift $\Delta\phi_0$ of the cavity field.}\label{fig:transmission}
\end{figure}

We will now discuss more precisely the nature of the phase instabilities we address in this work.  At optimal pulse compression timing, the phase of the drive signal $\phi_0$ is such that electrons at the center of the dispersed pulse arrive during when electric field oscillation in the cavity is $E(t)=0$.  This is achieved by varying the phase of the signal before high-power amplification. The momentum transferred to an electron pulse by an electric field $E(t)=E^z_0\exp\left({-i\omega_{\textup{d}} t +\phi_0}\right)$ is given by
\begin{equation}\label{eqn:momentumtransfer}
p_z = eE^z_0\int \exp\left({-i\omega_{\textup{d}} t +\phi_0}\right)\textup{d}t,
\end{equation}
where $E^z_0$ is the cavity field along the propagation axis of the electron pulse, $\omega_{\textup{d}}=2\pi f_{\textup{d}}$ and the integral spans the time during which an electron pulse interacts the cavity field.  By Eqn.~\eqref{eqn:momentumtransfer}, phase fluctuations, which we denote as $\Delta\phi_0$, cause variations in the average momentum transferred to electrons in the pulse $\Delta p_z$ (see Fig.~\ref{fig:transmission} a)) and consequently yield a change in the arrival time at the temporal focus of the cavity given by $\Delta t=-\Delta\phi_0/\omega_{\textup{d}}$~\cite{Pasmans2013}.  External phase fluctuations arise primarily from two sources.  The first is phase instabilities caused by high-power amplification which we write as $\Delta\phi_0^{\textup{A}}$.  The second source of phase drift arises from frequency detuning between the cavity resonance and the drive signal, $\omega_{\textup{d}}-\omega_0$. This leads to changes in the microwave transmission properties of the cavity (see Fig. 1 b)). In the vicinity of resonance ($\Delta \omega < \delta$), detuning produces a corresponding phase change in the cavity~\cite{Pasmans2013} given by
\begin{equation}\label{eqn:detuning}
\Delta\phi_0^{\textup{d}} = 2Q\frac{\omega_{\textup{d}}-\omega_0}{\omega_0}.
\end{equation}
Clearly, both $\omega_{\textup{d}}$ and $\omega_{\textup{0}}$ impact Eqn.~\eqref{eqn:detuning} and can vary independently. $\omega_{\textup{d}}$ is determined by the value of the oscillator repetition rate and thus variations on the order of $0.1-1~$kHz are expected on timescales on the order of days. The cavity resonance is sensitively a function of temperature~\cite{Pasmans2013}, for which changes in the range of a few mK yield variations in $\omega_{\textup{0}}$ also on the order of $0.1-1~$kHz.  These sources of detuning yield timing changes in excess of $\Delta t = \Delta\phi_0^{\textup{d}}/\omega_{\textup{d}} > 100~$fs.  We write the total arrival time drift due to both sources of phase fluctuations as $\Delta t =-\frac{1}{\omega_{\textup{d}}}\left(\Delta\phi_0^{\textup{A}}+\Delta\phi_0^{\textup{d}}\right)$ which has the effect of shifting the pump-probe delay (time-zero) during the course of an experiment, thus degrading the temporal resolution of the instrument.  It is therefore critical to accurately sample the electric field and phase changes occurring \it in \rm the resonant cavity to determine the relevant timing fluctuations impacting electron pulses in a UED instrument.

Detection of phase changes is achieved by measuring the signal transmitted \it through \rm the cavity using an integrated pick-up antenna (see Fig. 1) which retrieves the electric field inside the cavity.  The collected signal is attenuated and passed through a saturated low-phase noise amplifier (Holzworth HX2400) to maintain a stable power of $+$12.0~dBm.  This signal is then sent to a phase detector (Holzworth HX3400).  The detector compares the phase of this signal (RF), $\phi_{\textup{cav}}$, with the phase of the pre-compression reference signal (LO), $\phi_{\textup{ref}}$.  Since both signals have the same frequency $f_{\textup{d}}$, the detector produces a DC voltage proportional the phase difference between the two signals $V^0_{\phi} = \kappa_{\textup{det}}\left(\phi_{\textup{ref}}-\phi_{\textup{cav}}\right)$. $\kappa_{\textup{det}}=320~$mV/rad is the phase detector constant determined via calibration. Due to the character of the phase response shown in Fig.~\ref{fig:transmission} b) we assume that $\Delta\phi_0$ arises dominantly from the compression cavity with secondary high-frequency components from the solid-state power amplifier, therefore, $\phi_{\textup{cav}}\rightarrow\phi_{\textup{cav}}+\Delta\phi_0$ and $V_{\phi}=V^0_{\phi}+\Delta V$. The changes in detector output voltage $\Delta V$ allow for phase drift to be monitored with respect to the initial starting value $V_{\phi}^0$ (i.e. optimal compression settings). Information from the phase detector signal may also be used to correct for phase drifts by precisely shifting the phase of the microwave drive signal. This ensures that phase changes relative to the initial set-point value are minimized.  We implement this using a low-noise fully analog microwave phase shifter (Analog Devices HMC928LP5E). The phase detector output voltage is amplified by a factor of 10 to optimize the detection resolution of a fully analog PID controller (Stanford Research Systems SIM 960) which has an operating bandwidth of 1~kHz.  A schematic of the active phase feedback system is illustrated in Fig.~\ref{fig:feedback_cavity}.
\begin{figure}
  \centering
    \includegraphics[width=8.5cm]{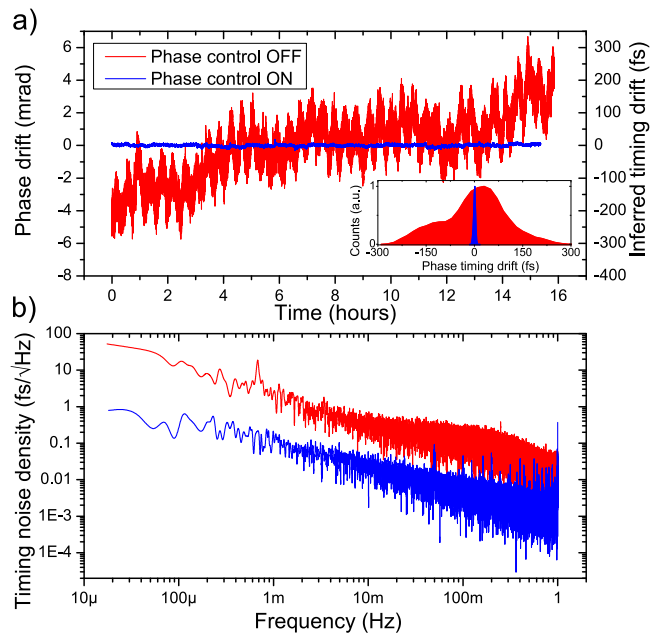}
  \caption{a) Electronic phase timing measurement of the detection system in Fig.~\ref{fig:feedback_cavity}.  Data for feedback off (red) and on (blue) are compared over the course of several hours and histograms of the phase distribution are shown in the inset. b) Additive timing noise power spectral density of the data shown in a).}\label{fig:PIDcomparison}
\end{figure}
We characterize the improvement of the instrument by measuring the detector output voltage over the duration of a typical experiment (several hours).  The output voltage is converted to phase and timing drift using $\kappa_{\textup{det}}$ and the drive frequency $\omega_{\textup{d}}$.  A comparison with the feedback and control on and off is presented in Fig. 3 a). Without feedback, detuning instabilities produce synchronization drift on the order of hundreds of fs (see Fig.~\ref{fig:PIDcomparison}~a)). With feedback, we see a significant improvement for which the synchronization drift below is 5~fs RMS.  The phase stabilization system reduces the power spectral density of the timing noise (relative to the signal pre-amplification) by over an order of magnitude across 5 low-frequency decades, as seen in Fig.~\ref{fig:PIDcomparison}~b). Both low-frequency thermal detuning drifts in the cavity and mid-frequency jitter are continuously corrected over long timescales.

\begin{figure}
  \centering
    \includegraphics[width=8.5cm]{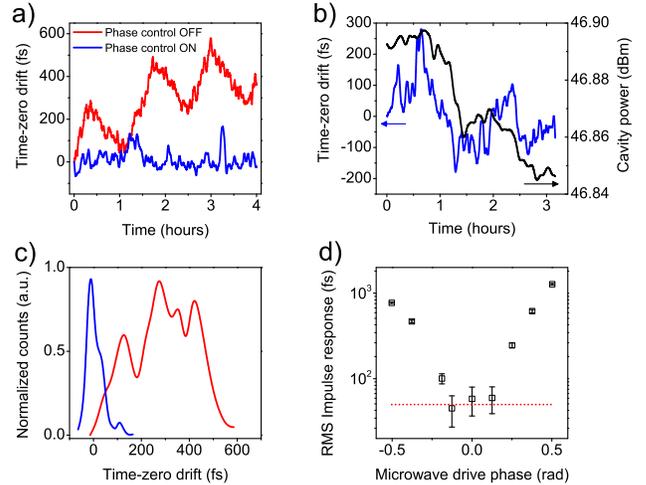}
  \caption{a) Electron pulse arrival time measured with phase feedback control both off (red) and on (blue). b) Correlation between arrival time and cavity power arising from amplitude-phase error in the feedback system. c) Histograms of the traces shown in a) depicting the long-term stability of time-zero. d) The temporal impulse response function of the UED instrument as a function of phase for a bunch charge of 0.2~pC and cavity power of 44.7~W. The dashed red line represents the time-resolution of the streak camera.}\label{fig:streakcomparison}
\end{figure}

We also directly measure the change in the arrival time of the electron pulses with respect to a femtosecond optical pump pulse using an optically triggered 10~GHz streak camera similar to that presented in Ref.~\cite{KassierRSI2010}.  By tracking the center of the streaked electron pulse on a CCD camera, we may quantify timing drifts in a manner which replicates a typical pump-probe UED experiment up to the limits imposed by the streak camera itself.  We operate at a laser amplifier repetition rate of 1~kHz and a CCD exposure time of 1~s. The streaked electron pulse it fit to a 2D Gaussian function to accurately determine the center position.  A relationship between the streaked pulse position and time depends on the geometry of instrument and also the circuit properties of the streak camera.  A calibration measurement was performed and a streak velocity of $v_s=84~\upmu$m/ps was determined.  The streak camera itself has a finite temporal resolution which we determine to be $\tau_{s}\approx 50~$fs RMS using the statistics of an unstreaked electron pulse and the pulse spot size~\cite{Gao2013,KassierRSI2010}. A comparison of the electron pulse arrival time stability with and without feedback control is shown in Fig.~\ref{fig:streakcomparison}~a) and c).  We find an significant improvement in the arrival time stability to better than $\Delta t = 50~$fs over the duration of many hours. This result is a significant improvement for high-brightness multi-shot ultrafast electron diffraction systems implementing microwave pulse compression~\cite{ChatelainAPL2012,Gao2012,Gao2013}.  The remaining temporal drift may be explained in terms of feedback over-correction due to amplitude-phase conversion arising in the phase detector.  Amplitude fluctuations lead to changes in DC voltage which cannot be differentiated by the analog feedback control system.  We determine an amplitude coefficient $-54~$mV/dBm by varying the power of the RF (compression cavity) signal.  When converted to phase using the phase detector constant, $\kappa_{\textup{det}}$, we have a phase error coefficient of $-0.17~$rad/dBm.  Amplifier drift on the order of $\pm~0.02~$dBm will yield a phase timing changes of roughly 180~fs which explains correlation between arrival time drift and cavity power depicted in Fig.~\ref{fig:streakcomparison}~b).  This effect could potentially be minimized by using limiters and/or microwave power stabilization hardware.  For the case where sufficient power amplifier stability may not be achieved, the phase detection system may be used passively (no feedback control) to log phase drift, the amplitude converted portion of the detector signal may be easily removed by measuring power, and a time-stamping procedure can be performed~\cite{Gao2013,PhysRevSTABScoby2010}.
Finally we characterize the temporal impulse response function (IRF) of the instrument.  The temporal RMS pulse duration $\tau_e$ of the electron pulse is given by the deconvolution of a reference (unstreaked) electron spot and streaked spot.  This is written as $\tau_e=v^{-1}_s\sqrt{\sigma_{\textup{ref}}^2-\sigma_{s}^2}$, where $\sigma_{\textup{ref}}$ and $\sigma_{s}$ are determined from fitting the unstreaked and streaked electron spot respectively to a 2D Gaussian function. The streak velocity $v_s=84~\upmu$m/ps was determined by a measurement of the streak field ring-down. Pulse images are taken as a function of cavity phase and $\sigma_{\textup{ref}}$ and $\sigma_{s}$ are determined for each acquisition.  The impulse response function is determined for a pulse charge of $0.2~$pC  and a forward power of 44.7~W. At each phase value, 30 pictures at 1~s exposure and 1~kHz repetition rate time are collected and the average values $\langle\sigma_{\textup{ref}}\rangle$ and $\langle\sigma_{s}\rangle$ are determined. The averages reflect $3\times10^{4}$ individual electron pulses are limited by jitter faster than the 1~s CCD integration time. The IRF as a function of phase is shown in Fig.~\ref{fig:streakcomparison}~d).  We find that the IRF approaches 50~fs for optimal power and phase, at which point the measurement is limited by the temporal resolution of the streak camera.  We measure an upper limit IRF of $\tau_e=45\pm~2~$fs (106~fs full-width at half-maximum). We note that this is an improvement by over a factor of 3 when compared to previously reported instrument performance~\cite{ChatelainAPL2012,Zandi2017,Gao2012}.  We attribute this enhancement in performance to the direct generation of phase-locked microwaves, the use of stable continuous-wave high-power microwave amplification in conjunction with active phase stabilization.  The total temporal impulse response function $\tau$, is given by $\tau=\sqrt{\tau_e^2+\Delta t^2}$.  Including long-term drift shown in Fig.~\ref{fig:streakcomparison}~c) of $\Delta t=50~$fs, yielding $\tau\simeq 68~$fs.  To the best of our knowledge, this is the lowest measured temporal impulse response function for a high-brightness, multi-shot, sub-relativistic ultrafast electron diffraction instrument.

In conclusion, we have achieved ultrafast electron diffraction with pulse compression using phase-locked microwaves synthesized directly from a mode-locked oscillator. Furthermore, we have designed and characterized a high-performance integrated microwave phase feedback system which compensates for detuning induced phase fluctuations in compression cavities improving the laser-microwave synchronization level by a factor of 10 to below 5~fs RMS.  The long-term arrival time stability of the electron pulse is also improved from $>200~$fs to $<50~$fs RMS with phase stabilization.  The impulse response function of the instrument is measured to be less than 70~fs RMS over many hours.  Our results illustrate that high-brightness UED instruments can be realized in a simpler approach and pushed to a regime where time-resolution is limited by the duration of the compressed electron pulse, not synchronization quality.

\begin{acknowledgments}
This work was supported by the Natural Sciences and Engineering Research Council of Canada (NSERC), the Fonds de Recherche du Qu\'ebec - Nature et Technologies (FRQNT), the Canada Foundation for Innovation (CFI) and Canada Research Chairs (CRC) program.
\end{acknowledgments}

\bibliographystyle{unsrt}
\bibliography{bibliography}

\end{document}